\documentclass[useAMS,usenatbib]{mn2e}

\usepackage{times,epsfig,natbib,amssymb,amsmath,graphics,longtable,threeparttable}
\usepackage{rotating}

\def\sun{\hbox{$\odot$}}
\def\cm3{cm$^{-3}$}

\def\lsun{L$_{\odot}$}
\def\rsun{R$_{\odot}$}
\def\mdot{$\dot{\rm M}$}

\def\msun{M$_{\odot}$}

\def\two{\ts {\,\sc ii}}

\def\beq{\begin{equation}}
\def\eeq{\end{equation}}

\def\lesssim{\mathrel{\hbox{\rlap{\hbox{\lower4pt\hbox{$\sim$}}}\hbox{$<$}}}}
\def\gtrsim{\mathrel{\hbox{\rlap{\hbox{\lower4pt\hbox{$\sim$}}}\hbox{$>$}}}}

\def\lesssim{\mathrel{\hbox{\rlap{\hbox{\lower4pt\hbox{$\sim$}}}\hbox{$<$}}}}
\def\gtrsim{\mathrel{\hbox{\rlap{\hbox{\lower4pt\hbox{$\sim$}}}\hbox{$>$}}}}

\def\ha{H$\alpha$}
\def\hbeta{H$\beta$}

\def\two{{\,\sc ii}}

\def \lsun{\ifmmode{{\rm\ L}_\odot}\else{${\rm\ L}_\odot $}\fi}
\def \msun{\ifmmode{{\rm\ M}_\odot}\else{${\rm\ M}_\odot$}\fi}
\def \rsun{\ifmmode{{\rm\ R}_\odot}\else{${\rm\ R}_\odot$}\fi}

\newcommand{\kms}{km s$^{-1}$}                         
\def \mdot{\ifmmode{{\rm\dot{M}}}\else{${\rm\dot{M}}$}\fi}

\newcommand{\apjl}{ApJ Let.}

\newcommand{\apjs}{ApJS}
\newcommand{\aap}{A\&A}

\def\cmfgen{{\sc cmfgen}}

\newcommand{\hii}{H\,{\sc ii}{}}

\title[Blue shifted emission peaks in SNe~II]{Analysis of blue-shifted emission 
peaks in type II supernovae}
\author[Anderson, Dessart et al.]
{J.~P. Anderson$^{1,}$$^{2}$\thanks{E-mail:
janderso@eso.org},
L. Dessart$^{3}$, 
C.~P. Gutierrez$^{4,2}$,
M. Hamuy$^{2,4}$,
N.~I. Morrell$^{5}$,
M. Phillips$^{5}$,
\newauthor
G. Folatelli$^{6}$,
M.~D. Stritzinger$^{7}$,
W.~L. Freedman$^{8}$,
S. Gonz\'alez-Gait\'an$^{2}$,
\newauthor
P. McCarthy$^{8}$,
N. Suntzeff$^{9}$,
J. Thomas-Osip$^{5}$\\
$^{1}$European Southern Observatory, Alonso de Cordova 3107, Vitacura, Santiago, Chile\\
$^{2}$Departamento de Astronom\'ia, Universidad de Chile, Casilla 36-D, 
Santiago, Chile\\
$^{3}$Laboratoire Lagrange, UMR7293, Universit\'e Nice Sophia-Antipolis, CNRS,
Observatoire de la C\^{o}te d'Azur, 06300 Nice, France \\
$^{4}$Millennium Institute of Astrophysics, Casilla 36-D, 
Santiago, Chile\\
$^{5}$Carnegie Observatories, Las Campanas Observatory, Casilla 601, La Serena, Chile\\
$^{6}$Institute for the Physics and Mathematics of the Universe (IPMU), University of Tokyo, 5-1-5 Kashiwanoha, Kashiwa, Chiba 277-8583, Japan\\
$^{7}$Department of Physics and Astronomy, Aarhus University, Ny
Munkegade 120, DK-8000 Aarhus C, Denmark\\
$^{8}$Observatories of the Carnegie Institution for Science, Pasadena, CA 91101, USA\\
$^{9}$George P. and Cynthia Woods Mitchell Institute for
  Fundamental Physics and Astronomy, Department of Physics and Astronomy,
  Texas A\&M University,\\ College Station, TX 77843, USA\\
}

\begin{document}

\date{}

\pagerange{\pageref{firstpage}--\pageref{lastpage}} \pubyear{2012}

\maketitle

\label{firstpage}

\begin{abstract}
In classical P-Cygni profiles, theory predicts emission to peak at
zero rest velocity.
However, supernova spectra exhibit emission 
that is generally blue shifted.
While this characteristic has been reported in many supernovae, it is rarely
discussed in any detail. 
Here we present an analysis of \ha\ emission-peaks
using a dataset of 95 type II supernovae, quantifying their strength and time evolution.
Using a post-explosion time of 30\,d, we observe a systematic blueshift
of \ha\ emission, with a mean value of --2000\,\kms.
This offset is greatest at early times but vanishes as supernovae become nebular.
Simulations of Dessart et al. (2013) match the 
observed behaviour, reproducing both its strength and evolution in time.
Such blueshifts are a fundamental feature of supernova spectra as
they are intimately tied to the density distribution of ejecta, 
which falls more rapidly than in stellar winds.
This steeper density structure causes line emission/absorption to be much more confined;
it also exacerbates the occultation of the receding part of the ejecta, biasing line emission 
to the blue for a distant observer.
We conclude that blue-shifted emission-peak offsets of several thousand \kms\ are a generic property of
observations, confirmed by models, of photospheric-phase type II supernovae.
\end{abstract}


\section{Introduction}
Type II supernovae (SNe~II henceforth) are classified through the presence
of hydrogen Balmer lines in their spectra (see \citealt{min41} for initial spectral
classification of SNe, and \citealt{fil97} for a review). 
Spectral line formation occurring in optically thick, rapidly expanding ejecta produce
spectral features to appear with a P-Cygni profile
morphology.
An observed P-Cygni profile is characterised by an absorption feature which is blueshifted,
and an emission feature found closer to zero rest velocity.
The blue-shifted absorption feature occurs due to material moving towards the observer which 
obscures continuum emission in the line of sight. As line
opacity acts in addition to continuum opacity,
the velocity at maximum absorption gives an estimate of the
photospheric velocity.
In classical P-Cygni theory (see e.g. \citealt{sob60,cas70}), 
the profile emission peaks at the rest wavelength of the corresponding line, i.e.,
at zero Doppler velocity in the rest frame. 
However, as first noted in the early-time spectra of SN~1987A \citep{men87}, 
observations reveal emission peaks that are often blue shifted, by as much
as several thousand \kms\ (also note that such blueshifts were briefly mentioned
in \citealt{che76}).
To explain this property in SN~1987A, \cite{chu88} presented a modified version 
of the standard model for line formation which allowed for diffuse reflection of resonance 
radiation by the SN photosphere. 
This observation was further outlined by \cite{elm03_2} for the case of the well-observed
type II-Plateau (II-P) SN~1999em, while similar blue-shifted features have been discussed for
SN~1988A \citep{tur93}, and SN~1990K \citep{cap95}. More recently, this
observation was reported for the sub-luminous type II-P SN~2009md \citep{fra09}.\\
\indent While the above summary shows that blue-shifted emission
peaks have been observed and documented in the past, it still appears to be a
relatively unexplored feature in the SN community, especially given
the large number of individual SN~II studies which have now been published. Indeed, there
are many cases where such blueshifts are present, and seen to evolve in
published spectral sequences; yet, little or no discussion is found in the
respective papers (see e.g. \citealt{ham01,leo02_2,leo02,bos13}).\\
\indent From the modeling point of view \citet{des05_2} presented a study of 
the formation of P-Cygni line profiles in
SNe~II based on non-Local Thermodynamic Equilibrium (LTE) steady-state radiative transfer models. 
In contrast 
to \citet{chu88}, these authors concluded that the emission-peak blueshift
is intimately related to the steep density fall-off of SN ejecta, causing
strong occultation and optical-depth effects. 
Since the work of  \citeauthor{des05_2}, \cmfgen\
has been augmented to incorporate
time dependence in the radiative-transfer equation and in the statistical-equilibrium equations,
allowing a full time-dependent solution for the SN radiation based on more physical models of the
progenitor star and its terminal explosion \citep{des08_2,des10_2,hil12}.
As discussed below, the basic conclusions and predictions on P-Cygni 
profile formation made by \citeauthor{des05_2} are confirmed by recent 
models of type II SNe that include these upgrades \citep{des13}. The comparison between such models
and observed SN spectral features nourishes the discussion presented in this paper.\\
\indent We present a systematic analysis of the \ha\ emission peak
wavelengths for SN~II, and discuss these observations in terms of theoretical understanding
and modeling of spectral line formation. We do this to bring attention to the frequency of such
observations, and to outline how they may be used to foster a better
understanding of SN~II progenitor and explosion physics.\\
\indent The paper is organized as follows. In the next Section the data
sample used for analysis is discussed, together with a brief overview of reduction and
analysis processes. Then in \S\ 3 our results on the
distributions and evolution of \ha\ P-Cygni emission peak velocities are presented. In
\S\ 4, we outline the theoretical framework for understanding such features, and use numerical simulations 
to explain their measurement and observation.
The physical origin of these observations is discussed in \S\ 5, where 
we also connect this property to other SN characteristics.
Finally, we present our conclusions in \S\ 6.

\begin{figure}
\includegraphics[width=8.5cm]{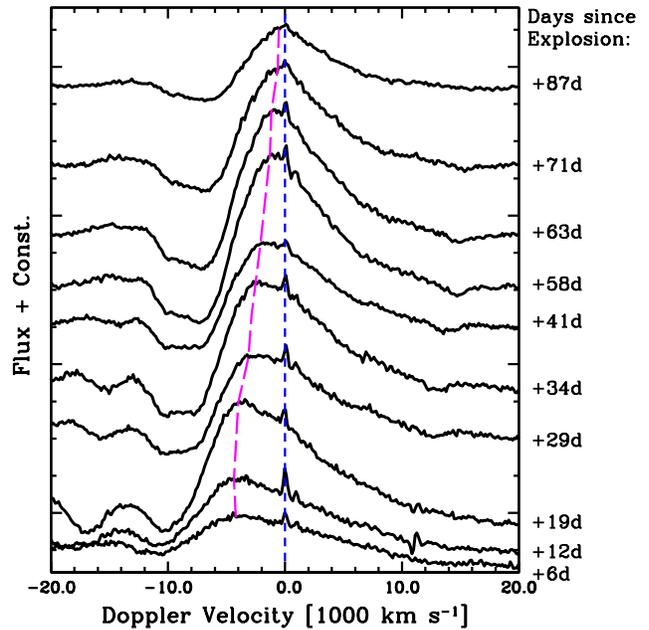}
\caption{\ha\ spectral sequence of SN~2007X. The rest velocity of the SN
position, as measured from the host \hii\ region narrow \ha\ emission, is
indicated by the short dashed blue line (i.e. at zero velocity). The long
dashed
magenta line shows the evolution of the Doppler velocity of the emission peak.
\label{fig07X}}
\end{figure}

\begin{figure}
\includegraphics[width=8.5cm]{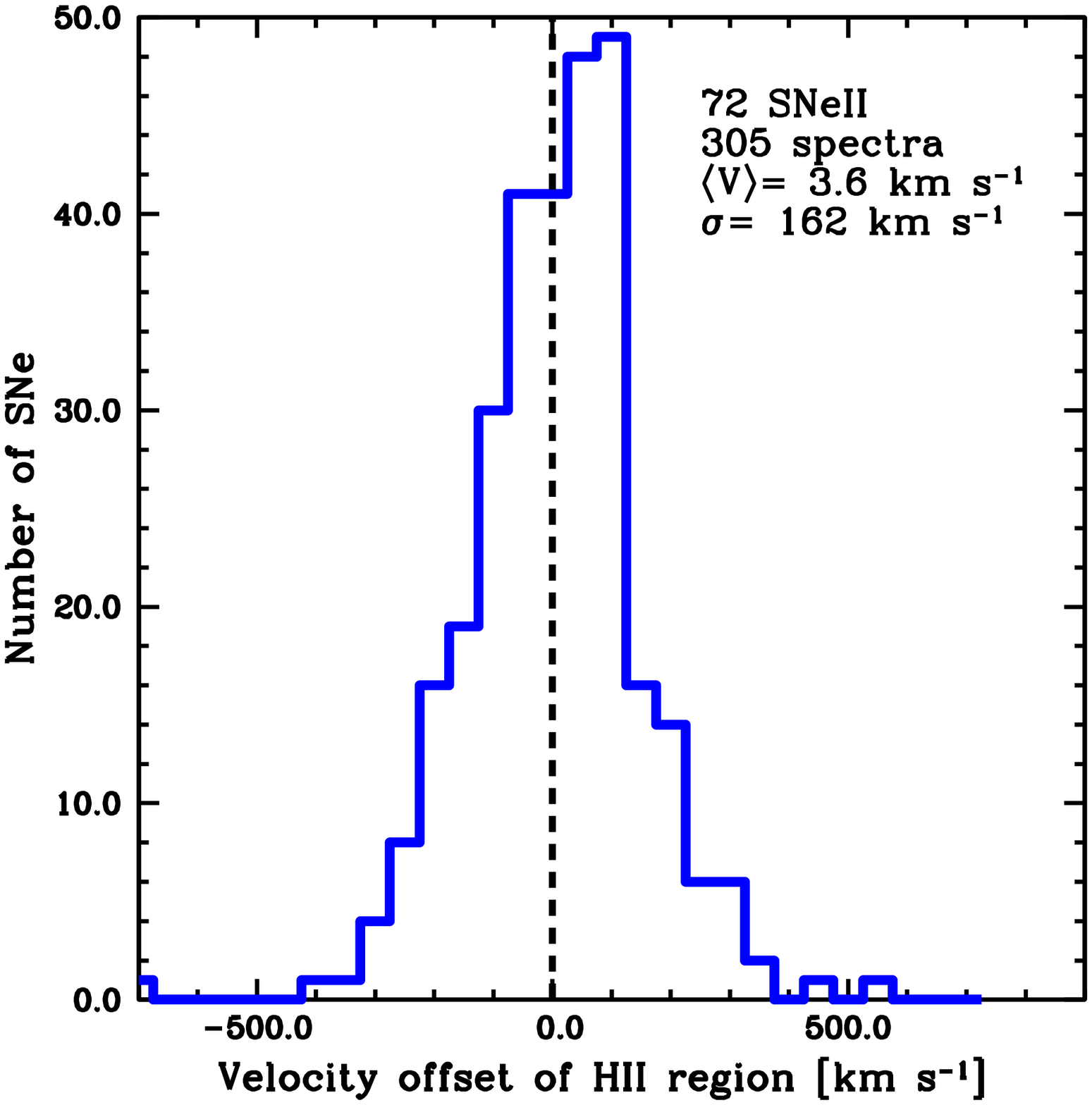}
\includegraphics[width=8.5cm]{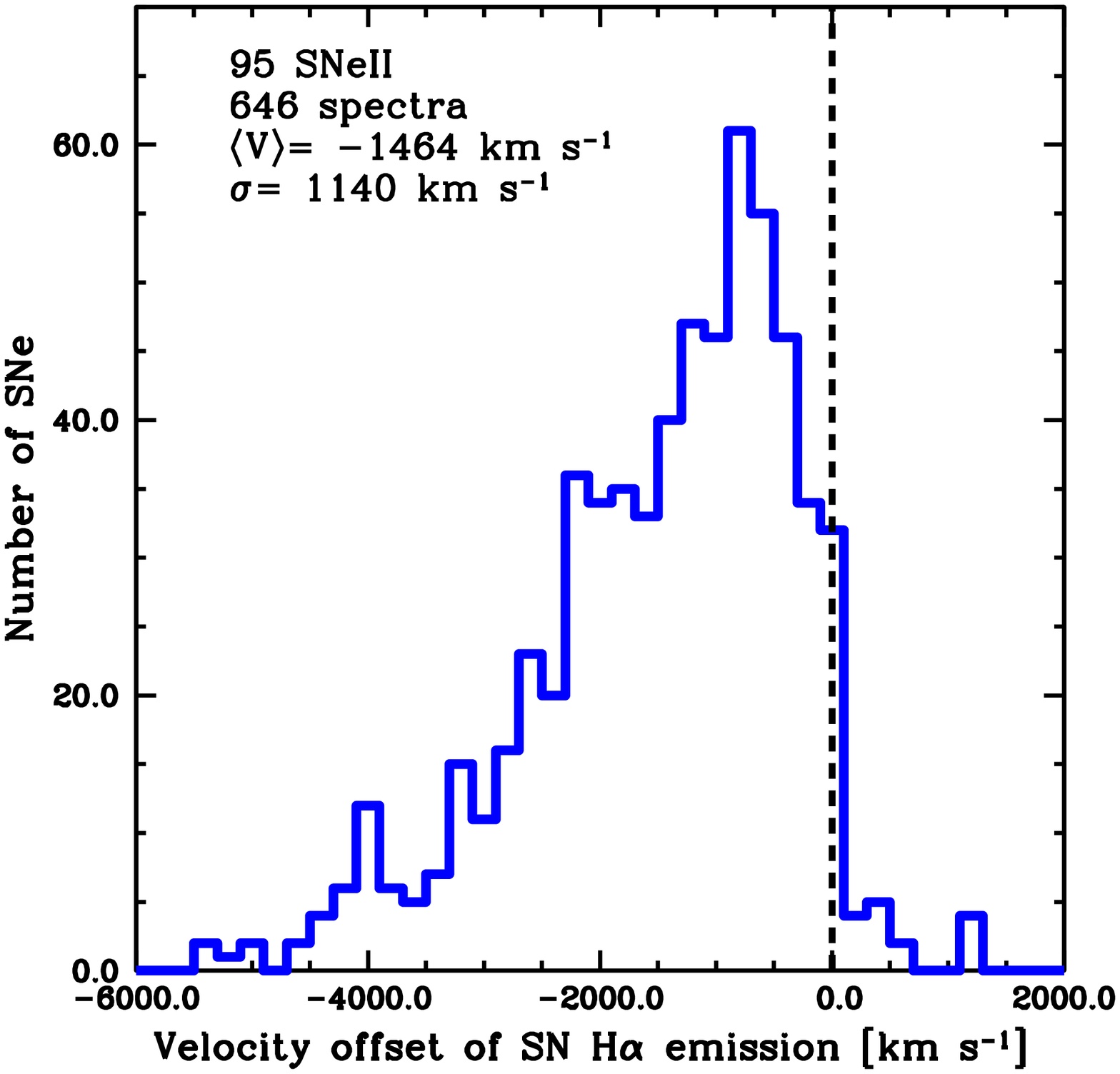}
\caption{\textit{Top:}  Histogram showing the velocity of the narrow peak emission (associated with 
a coincident \hii\ region) detected in each SN spectrum, and given with respect to the
host galaxy recession velocity. 
\textit{Bottom:}  Histogram of the distribution of velocity offsets of \ha\ SN emission for the entire sample of 
type II SN spectra (corresponding to different post-explosion epochs).}
\label{fighists}
\end{figure}

\begin{figure}
\includegraphics[width=8.5cm]{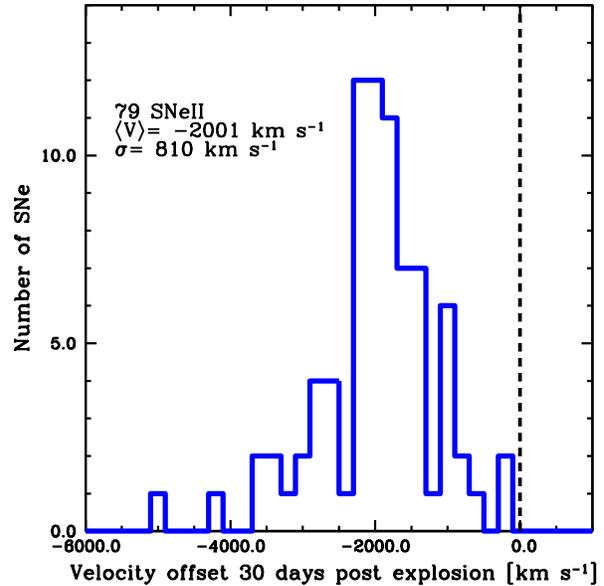}
\caption{Histogram of the distribution of SN \ha\ emission velocity offsets for observations at 30\,d after explosion.}
\label{fighist2}
\end{figure}

\section{Data and analysis methods}

The data sample was obtained through various SN follow-up campaigns. 
These are: 1) the
Cerro Tololo SN program (CT, PIs: Phillips \&\ Suntzeff, 1986-2003); 2) the
Cal\'an/Tololo SN program (PI: Hamuy 1989-1993); 3) the Optical and Infrared
Supernova Survey 
(SOIRS, PI: Hamuy, 1999-2000);
4) the Carnegie Type II Supernova Program (CATS, PI: Hamuy, 2002-2003); and 5) the
Carnegie Supernova Project (CSP, \citealt{ham06}, PIs: Phillips \&\ Hamuy,
2004-2009).
An initial analysis of the $V$-band light-curve morphologies of this sample
has recently been published in Anderson et al. (accepted), and we refer
the reader to that paper for in-depth description of SN light-curve parameter
measurements. 
In addition, an initial analysis of the spectral
diversity of SNe~II, concentrating on \ha\ profiles, was recently published in
Gutierrez et al. (accepted).
The data taken from the above follow-up programs used in the current analysis
amounts to 646 optical wavelength spectra of 95 SNe~II. 
From the above surveys, we excluded events classified as type IIn, type IIb, and SN~1987A-like. 
Furthermore, we only include in the current analysis SNe which have well-constrained explosion
epochs (see below), and more than two spectral observations during the photospheric phase.
We note that the follow-up surveys contributing to the current sample were
magnitude limited.\\
Spectra were reduced and extracted in the standard way, and the reader is referred to
\cite{fol13} for a detailed outline of the procedures employed (see also \citealt{ham06}). These are analogous to those 
applied to the CSP SN~Ia sample. 
In summary, 2d spectra were bias subtracted and flat fielded before 1d extraction
using the \textit{apall} task in IRAF. 1d spectra were then wavelength calibrated through observations of arc lamps,
and were finally flux calibrated using observations of spectrophotometric standard stars. 
The typical RMS of individual wavelength solutions less than 0.4 \AA, hence this
uncertainty brings negligible error into the velocity estimations.
The database of spectra were obtained with many different telescopes and instruments.
Typically the extracted 1d spectra have wavelength resolutions between 5--8 \AA, corresponding
to velocity resolutions of 230--370 \kms. The S/N ratio of spectra vary, depending on object brightness
and distance from the observer. In the majority of cases the S/N of the 1d spectra near to the 
\ha\ wavelength region of interest is more than 20.\\
\indent To proceed with measuring the wavelength (and therefore velocity) shift of
\ha\ emission peaks, we take host galaxy heliocentric recession
velocities from NED\footnote{http://ned.ipac.caltech.edu/}, and use 
these to place the observed spectra on to the rest wavelength
of the SNe. In addition, many SNe~II show the presence of narrow emission
lines in their spectra due to \hii\ regions at the SN site. When available, these 
give a more accurate Doppler velocity for the SN environment.
Hence, we measure the wavelength of these regions as a) a sanity check of host 
galaxy recession velocities, and b) to use these in place
of host galaxy recession velocities where measurements are possible (this distribution
is further discussed below, and is presented in Fig.\ref{fighists}).\\
\indent The emission peak wavelength of the SN
\ha\ profile (and the wavelength of the narrow host \hii\ region) in each
spectrum of the sequence for each object is measured. 
This is done by employing the \textit{splot} routine in IRAF\footnote{IRAF is distributed 
by the National Optical Astronomy Observatory, which is operated by the 
Association of Universities for Research in Astronomy (AURA) under 
cooperative agreement with the National Science Foundation.}. 
Within \textit{splot} the `k' measurement is used to fit a single Gaussian to the emission part of 
the P-Cygni profile. The presence of narrow \ha\ from an underlying HII region can complicate the
measurement of this peak. Therefore, in cases where strong \ha\ emission dominates over the
\ha\ from the SN we do not include measurements from those spectra. Where narrow \ha\ is present but
weaker, then this is removed from the spectra by simply tracing a line across the base of its flux.
Once the above has been taken into consideration, we fit the emission multiple times, changing the
wavelength range used, until a fit is found to be satisfactory `by-eye' (i.e. 
that the peak of the Gaussian coincides with the peak of the emission). Through this process
measurements are obtained using a wavelength window of around $\pm$ 50--100 \AA, around the
wavelength of the peak emission.
Once a peak wavelength is measured it is converted to a SN velocity using 
the inferred recession velocity, as described above.
When the host galaxy recession velocity is employed, this brings a velocity
error of $\approx$ 200\,\kms\ (in extreme cases) due to the fact that recession
velocities are measured at galaxy centres, while SNe explode at a range of
galacto-centric distances, and hence due to galaxy rotational velocities,
could have rest velocities offset from those published in the literature (note, this is 
a very conservative error estimate: see Fig.\ref{fighists}). In addition,
from our experience measuring peak emission wavelengths, any single spectral
measurement has an error of several hundred \kms\ (due to the difficulty in
defining the peak, together with the spectral resolution of each observation). 
Hence, we assume a conservative
error of 500\,\kms\ on all SN velocity measurements.\\
\indent An example spectral sequence is shown in Fig.~\ref{fig07X}, where there is clear evidence for a
significant blueshift of the SN \ha\ emission peak with respect to the narrow
line from a coincident \hii\ region. In addition, one observes a significant evolution 
of velocities with time, in particular of the emission-peak offset.\\
\indent Before continuing to our results, we note that while this work
concentrates on the observed wavelength/velocity of the \ha\ emission peak,
significant velocity offsets are also found for the emission peaks of
other spectral features. At early times one observes very similar strength
velocity offsets for the emission peak of \hbeta. However, after several
weeks post explosion, an accurate measurement becomes impossible because
of the increased effects of line blanketing, in particular the overlap with neighbouring
Fe\,\two\ lines. Hence, one cannot follow the evolution of
\hbeta\ emission velocity for more than a few weeks after explosion. 
Another complication is that some spectral features are absent at early times, and so only
appear when line blanketing and line overlap are strong.
This is for example a problem with the spectral feature at 4450\,\AA\ seen 
in the earliest spectra of SN~2006bp \citep{qui07} and SN~1999gi \citep{leo02_2}, which
indeed models predict to be He\,\two\,4686\,\AA\ but strongly blue shifted due to a very steep density
gradient in the photospheric regions at early times \citep{des08}.
\ha\ is by far the strongest line in emission in SN~II spectra, and is present at all
epochs. This line is thus ideally suited for the study presented here,
and we continue using \ha\ as the tracer of velocity-shifted emission features 
for the remainder of this paper.

\section{Results}

\begin{figure*}
\includegraphics[width=17cm]{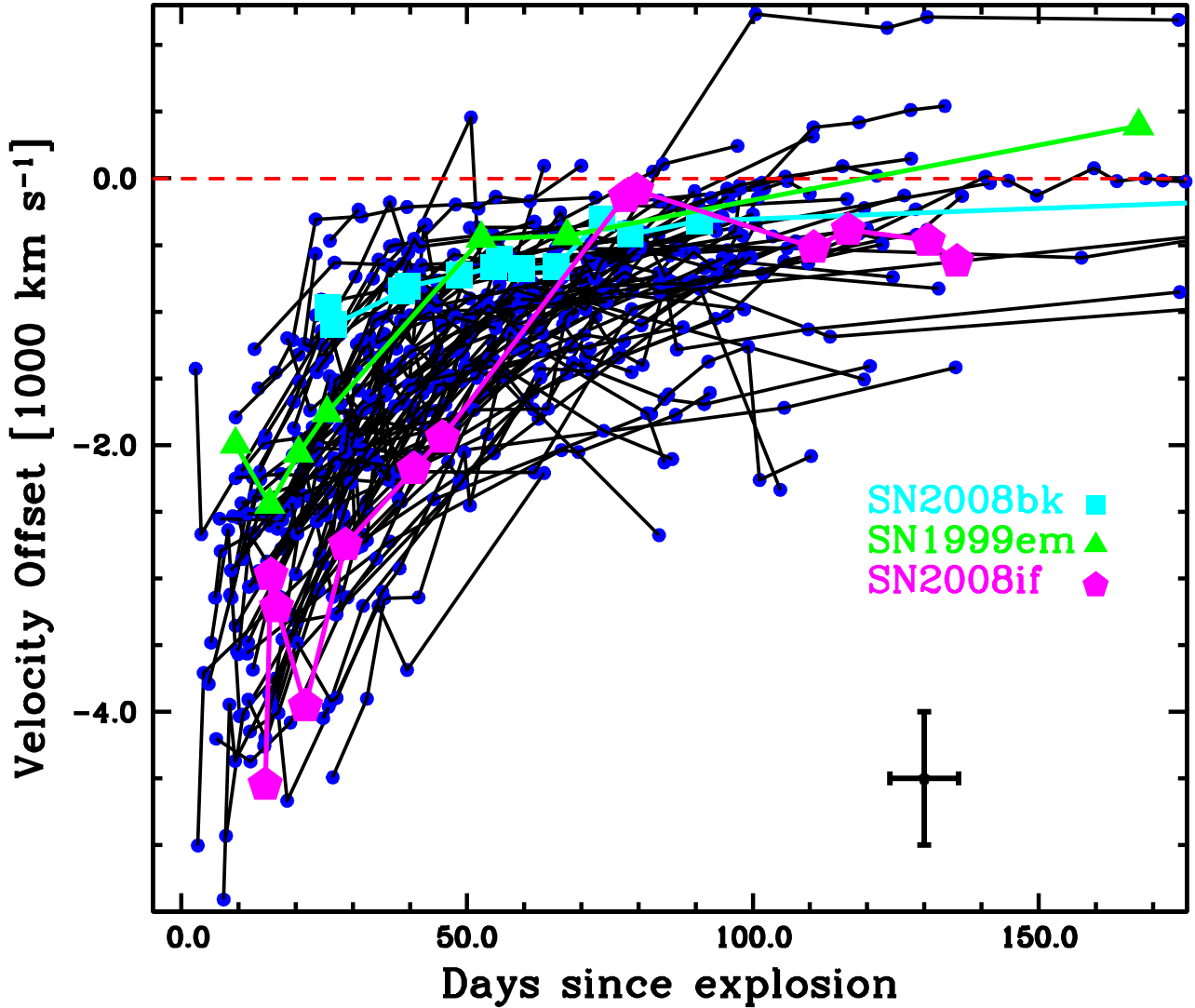}
\caption{ 
Evolution in time of the velocity offset of SN \ha\ emission peaks
for the 95 SNe~II in the current sample. Three events are shown: a sub-luminous event, SN~2008bk;
a prototype type II-P,
SN~1999em; and a faster declining event, SN~2008if. A standard error bar for all measurements is
given in black, where the origin of the velocity error is outlined in \S\ 2, and the time error 
is based on average errors of estimated explosion epochs (see Anderson et al. accepted, for a detailed description
of that process).}
\label{figevol}
\end{figure*}

In Fig.~\ref{fighists} (top panel), the distribution of velocities of narrow \ha\ emission
(associated with the coincident \hii\ region)
with respect to host galaxy recession velocities is presented. This
distribution contains 305 spectra for 72 SNe~II in our sample, and the mean 
velocity offset is 4\,\kms: i.e. essentially zero. The standard deviation of this distribution is
162\,\kms. These values match expectations, since published host galaxy
recession velocities will generally be velocities for the centre of each
galaxy, whereas SNe will be distributed throughout the galaxy and hence
will be found at velocities with a distribution centred on zero and a
standard deviation equal to expected rotational velocities of spiral
galaxies.\\
\indent In Fig.~\ref{fighists} (bottom panel), we present the velocity offset distribution of 
the (broad) SN \ha\ emission peaks for \textit{all} SNe
spectra within our sample. It is immediately apparent that the distribution is
heavily offset to significant negative velocities. Indeed, only 4\,\%\ of
events have positive velocities (red-shifted emission peaks). The mean velocity offset is 
--1464\,\kms\ with a standard deviation of 1140\,\kms.\\
\indent As noted above, when one observes spectral sequences of SNe~II, it is 
quite obvious that the blueshift of lines evolves significantly with
time. In Fig.~\ref{fig07X}, this evolution is shown for SN~2007X. 
To compare SNe in terms of absolute blue-shifted
velocities one needs to measure such a velocity at a consistent epoch. We adopt
a post-explosion time of 30\,d, where explosion epoch estimates are taken from
Anderson et al. (accepted), and interpolate measured velocities to this
time. This early epoch is used because it corresponds to times
when the diversity of blue-shifted velocities is high (see Fig.~\ref{figevol}). 
It also allows a significant fraction of the SN sample to be included, because numerous
SNe in our sample lack spectroscopic data prior to a month after explosion.
We note that the typical errors on our explosion epochs are $\pm$ 6 days (taken
from Anderson et al. accepted). The epochs used to interpolate velocities 
to 30\,d\, are generally within $\sim$ 5--15 days either side of this epoch.
The distribution of velocity offsets at 30\,d\ post explosion is presented in Fig.~\ref{fighist2}. 
The startling feature of this plot is that SNe have exclusive blue-shifted 
velocities at this epoch. Indeed, the mean of the distribution is
--2001\,\kms\ with a standard deviation of 810\,\kms. The other interesting
feature of the figure is the high velocity tail out to $\sim$\,4000 \kms. The 
overriding conclusion from the distributions presented above is that
significant blue-shifted emission velocities on the order of several thousand \kms\ are a
ubiquitous feature in SN~II spectra.\\
\indent After presenting the distribution of the velocity offsets of \ha\ emission
peaks at a representative post-explosion epoch, we now turn our attention to their evolution
with time. In Fig.~\ref{figevol}, this evolution is displayed for all 95 SNe
in our sample. While there is significant scatter at all epochs, a number of
interesting features are observed. Firstly, the highest velocities are
exclusively seen at early times. Very few SNe
show blue-shifted emission peaks higher than 2000\,\kms\ after 50\,d\ post
explosion, while at early times there is a significant number of SNe with
velocities in excess of 3500\,\kms. Secondly, the dispersion in velocity appears to decrease
significantly with time.\\
\indent In summary, significant blue-shifted velocities of several thousand \kms\ are a
very common feature of type II SN spectra. These velocities evolve in time quite
uniformly between SNe, eventually nearing zero as they progress to the end of the
photospheric phase.
In the following Section we discuss model spectra known to match the basic properties of SNe II-P
\citep{des13}, as well as the theoretical framework for understanding the origin of emission-peak
blueshifts.

\begin{figure*}
\epsfig{file=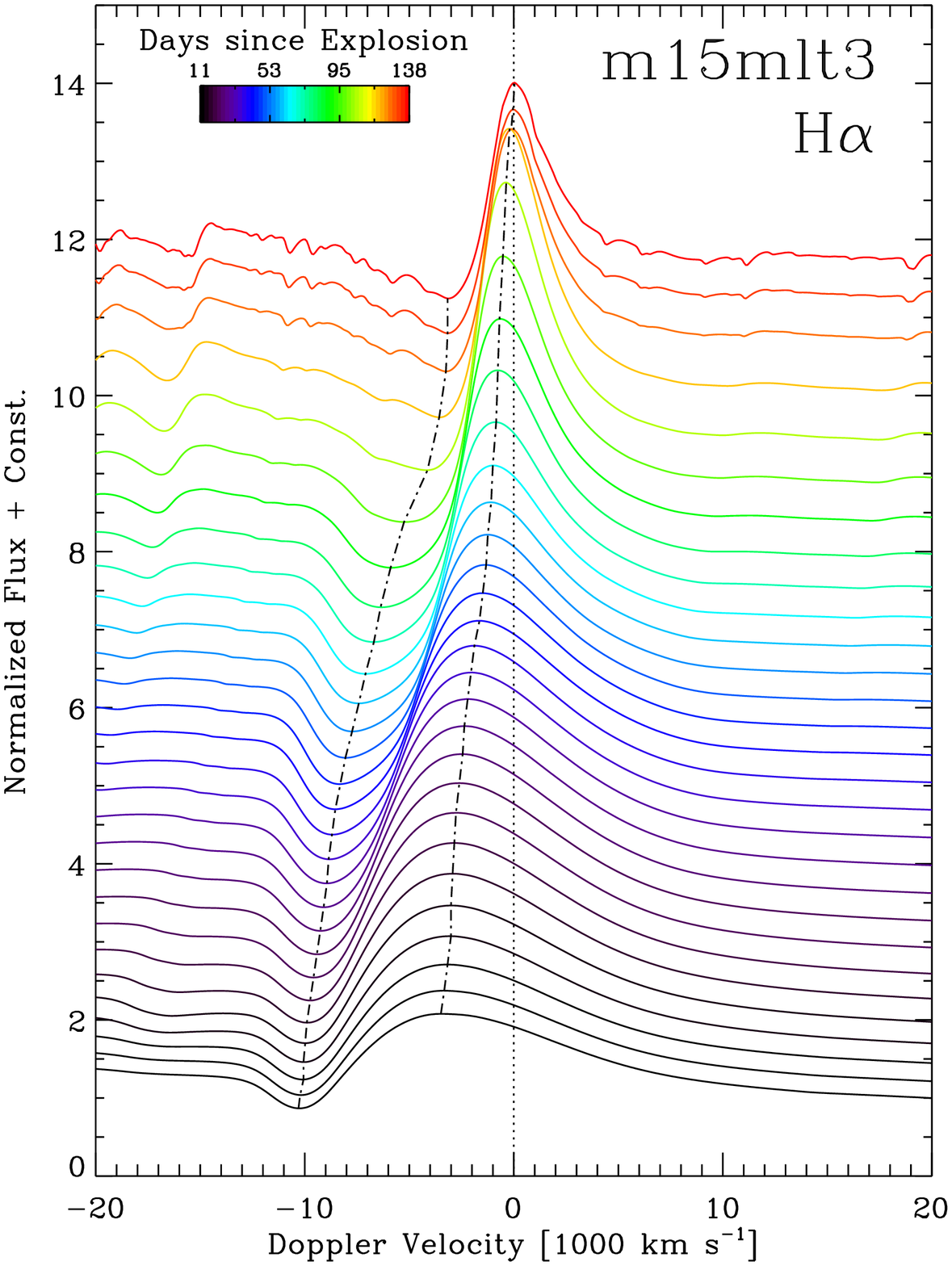,width=13cm}
\caption{Evolution of the H$\alpha$ spectral region from 11\,d (bottom curve) until
138\,d (top curve) after explosion in the SN II-P model m15mlt3 \citep{des13}.
The abscissa is the Doppler velocity with respect to \ha.
Individual times are color coded, and the time difference between consecutive models is 10\.\% of
the current time.
The two broken lines track the location of maximum absorption and peak emission in H$\alpha$.
The peak blueshift, strong at early times, decreases as the spectrum formation region recedes,
and eventually becomes zero at the end of the plateau phase (in this model at $\sim$\,140\,d).
\label{fig_prof_evol}}
\end{figure*}

\section{Insights from radiative-transfer simulations}

\subsection{Previous studies with \cmfgen}

As documented in the previous Section, the offset of P-Cygni profile emission to the blue is a
ubiquitous feature of all SNe~II observed during the photospheric phase.
In this Section, we investigate whether this generic property is also present in 
spectra produced by radiative-transfer simulations of SNe~II.\\
\indent Spectrum formation for SNe~II has been studied previously with \cmfgen. In
\cite{des05,des05_2}, steady-state non-LTE simulations
were used to discuss the physics of P-Cygni profile formation in early time spectra of SNe~1987A and 1999em.
The peak blueshift is reproduced for both events prior to the recombination phase, and a
detailed discussion of the origin of this property is given in \S\ 5 of \cite{des05}.
Later, in \cite{des08_2}, the discussion is extended to more advanced photospheric-phase epochs, when the photospheric
layers recombine, using a time-dependent solver for the  non-LTE rate equations.
This study focused primarily on the importance of time-dependent ionization,
and in particular its role for producing a strong H$\alpha$ line at the recombination epoch ---
little emphasis was put on the peak blueshift.\\
\indent In \cite{des13}, the radiative transfer in \cmfgen\ was improved by combining non-LTE and time
dependence (for the non-LTE rate equations and the moments of the
radiative-transfer equation). In addition,  the computation
is then performed on physically-consistent (although piston-driven) hydrodynamical explosions 
(see \citealt{hil12} for details and \citealt{des10_2} for an application to SN\,1987A).
Moreover, these simulations cover a range of progenitor evolution and explosion properties
for a 15\,\msun\ main-sequence star model and therefore allow one to inspect a number of dependencies
on SN~II-P radiation properties.\\
\indent In the present paper, we inspect the simulations of \cite{des13}, with a special focus on
the evolution of line profile morphology from early photospheric epochs to the
onset of the nebular phase. We focus on a representative sample of models,
namely s15N, m15r1, m15r2, m15os m15mlt1, m15mlt3, m15, m15e3p0, m15e0p6,
and m15Mdot (see  \citealt{des13} for details; these models cover a range of
properties for the same main-sequence star mass but different initial rotation rate, mixing-length parameter,
core overshooting prescription, or explosion energy).  
We also include model m15mlt1x3,
which is evolved the same way as m15 but with a mixing-length-parameter of 1.5 and
a mass loss enhanced by a factor of three compared to the standard red-supergiant (RSG) mass
loss rates provided by the `DUTCH' recipe in {\sc mesa}.
This produces a low envelope-mass RSG at the time of explosion, which, when exploded to yield
a 1.2\,B ejecta kinetic energy, yields a Type II-Linear light-curve morphology (see Hillier et al., in prep.).

\subsection{Results for SN~II-P simulations}
\label{sect_res_mod}

The first important result is that the peak-emission blueshift, 
observed in spectra of SNe~II (see, e.g., Fig.~1)
is also predicted by simulations at all times prior to the onset of the nebular phase.  
Figure~\ref{fig_prof_evol} shows the evolution of the H$\alpha$ region from 11\,d until 138\,d 
after explosion in model m15mlt3.
This model matches well the spectral and light curve evolution of SN\,1999em, except
for a prolonged plateau phase of $\lesssim$150\,d, which stems for the underestimated RSG
mass loss in the {\sc mesa} model (see \citealt{des13} for details).
The evolution of the velocity offset of peak emission agrees qualitatively and quantitatively
with the observations. Importantly, all the simulations presented in \citet{des13}
show the same behaviour, in agreement with observations (Fig.~6). Typical peak blueshifts are
on the order of $-$3200\,\kms\ at 10\,d after explosion and steadily decrease in strength
to eventually reach zero at the end of the plateau, as observed.\\

\begin{figure*}
\epsfig{file=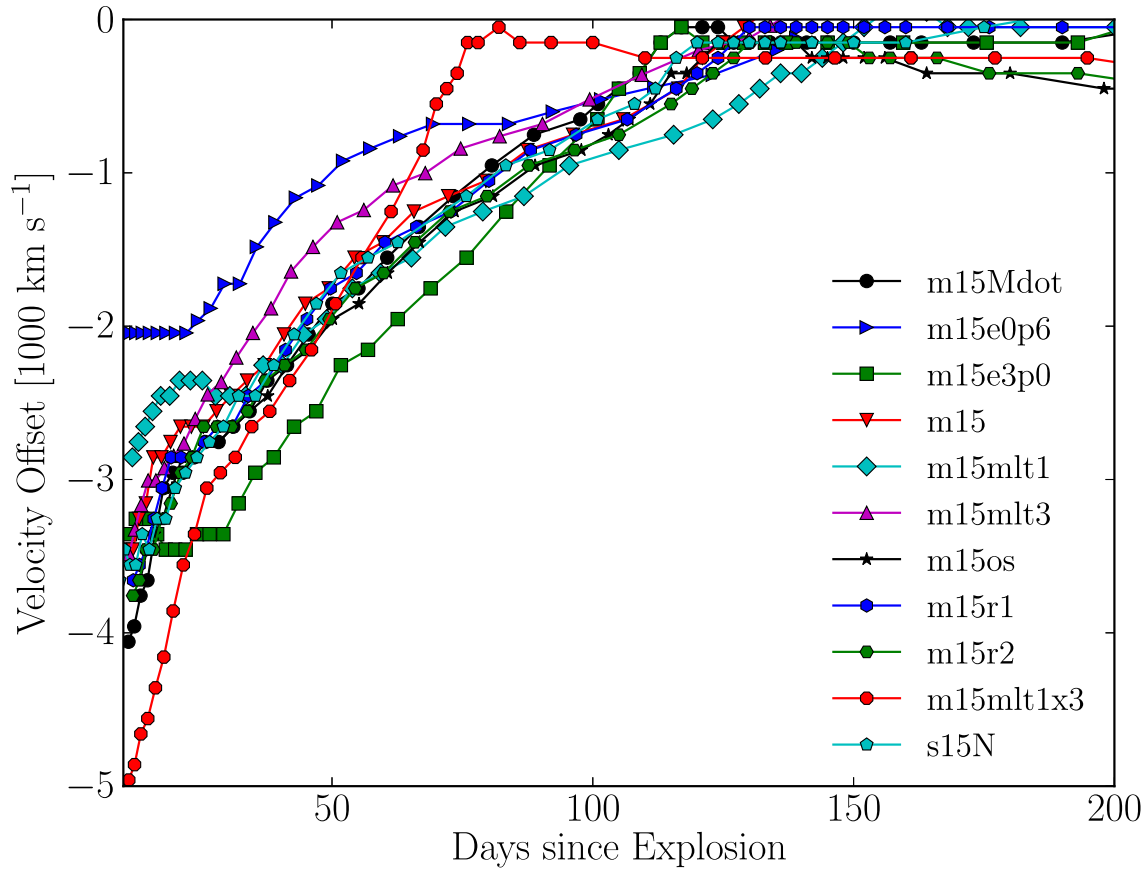,width=17cm}
\caption{Evolution of the velocity offset of H$\alpha$ peak emission with respect to rest wavelength
for the large set of models presented in \citet{des13}. All models follow the same trajectory,
except for models m15e3p0/m15e0p6 (3 and 0.6\,B, respectively), which have a larger/lower ejecta
energy than other models (i.e., 1.2\,B), and for model m15mlt1x3, in which the RSG mass loss
was artificially increased by a factor of 3. [See text for discussion].
\label{fig_path}
  }
\end{figure*}

\begin{figure*}
\epsfig{file=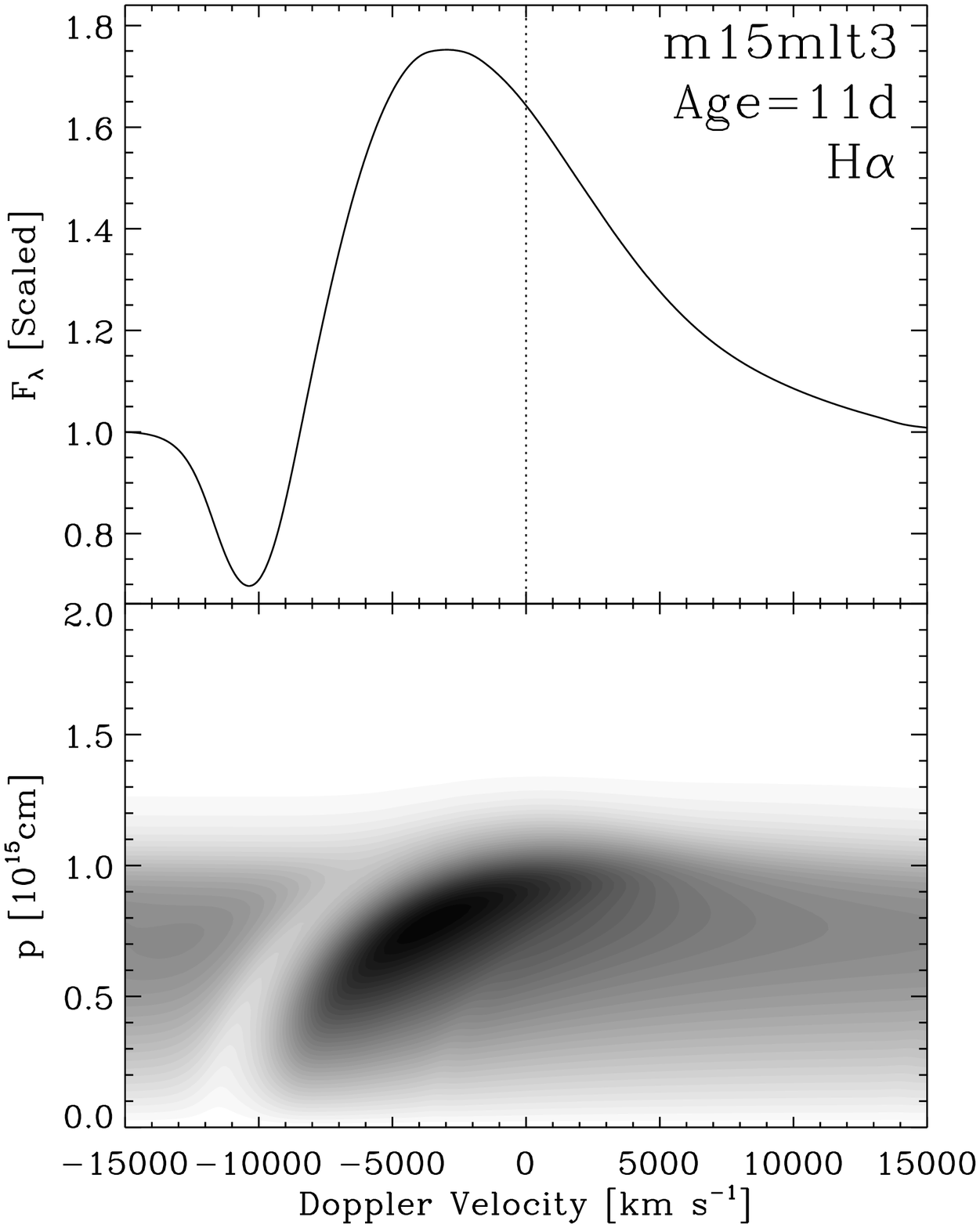,width=8.15cm}
\epsfig{file=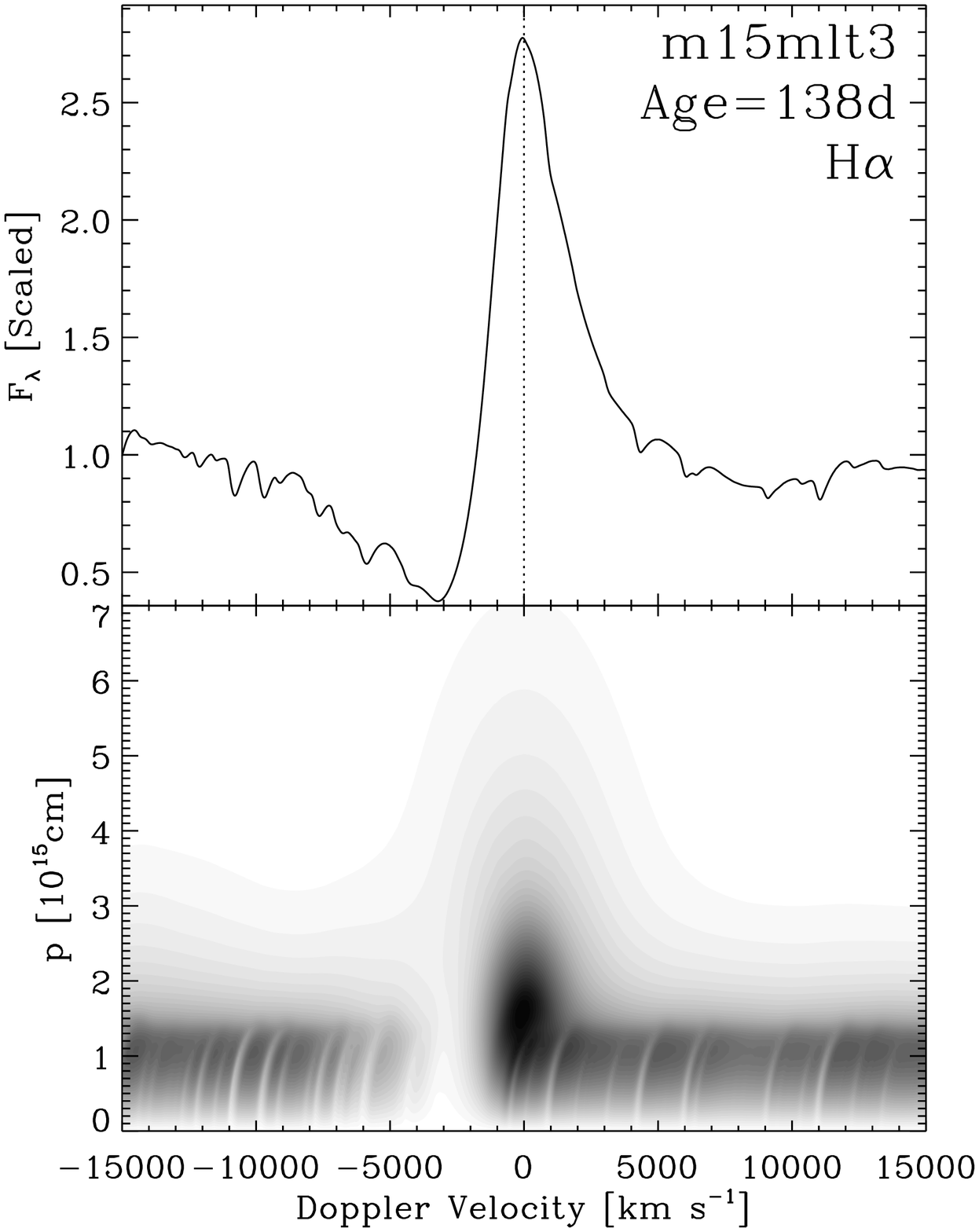,width=8.15cm}
\caption{Distribution of the quantity $p \times I(p)$, as seen by a distant observer at rest,
versus Doppler velocity $v$ and impact parameter $p$ (lower panels). When integrated over $p$, 
this gives the flux at $v$ (upper panels). 
Here, we show a SN II-P model at 11\,d (left) and 138\,d (right) after explosion.
The grayscale maps thus reveal the origin of the emergent radiation in different profile regions
(Doppler velocity) along rays with different $p$.
The main reason for the blueshift is the confinement of emission
within the column of material falling onto the `photo disk'. This disk is limited on the plane of the sky
by an impact parameter $p$ on the order of $R_{\rm phot}$. Because of obscuration effects,
and the lack of line emission at large impact parameters, the emission comes primarily from regions
that are blue shifted, and this naturally leads to peak-emission blueshift.
At late times (right panel), obscuration effects are still present, but there is now significant emission at
large impact parameters, which occurs symmetrically from approaching and receding regions.
In fact, photon emission is more extended at all wavelengths, stemming both the former He core and the
H-rich envelope.\label{fig_pip}}
\end{figure*}

\section{Discussion}

In previous sections it has been shown that significant blue-shifted velocities of \ha\ emission peaks are a common
feature of both observations \textit{and} models of SNe~II. Here we first
discuss the physical origin of these features and their diversity, before presenting two correlations of this property with other 
SNe~II transient measurements.

\subsection{The physical origin of blue-shifted emission peaks}

In contrast to \cite{chu88}, the origin of peak-emission blueshift 
seems to stem fundamentally from the steep density
profile that characterizes SN ejecta layers, which were originally part of the H-rich envelope. In our simulations,
this density distribution is well represented above $\sim$\,2000\,\kms\ (which is roughly where
the outer edge of the former He-core lies in 1.2\,B explosions of 15\,\msun\ RSG stars; \citealt{des10}) 
by a power law $\rho(v) \propto 1 /v^n$, with exponent $n$ on the order of 8.
As a consequence, line emission tends to be confined in space, and subject to strong
occultation effects for a distant observer. Instead of coming predominantly from the regions with large
impact parameter $p$ relative to the photospheric radius $R_{\rm phot}$ the bulk of the 
emission arises from rays with $p < R_{\rm phot}$, i.e. impacting the photo-disk limited
by $p = R_{\rm phot}$. As explained in \cite{des05_2}, the line source function
tends to exceed the continuum source function at the continuum photosphere, naturally
leading to line emission above the continuum flux level.
Because electron-scattering dominates the opacity, line photons have a relatively low
destruction probability and we can therefore see line emission from layers deeper than 
the continuum photosphere.\\
\indent Fig.~\ref{fig_pip} illustrates these effects for model m15mlt3 at 11\,d (left panel).
As time progresses, the situation evolves significantly for several reasons. The
spectrum formation recedes to deeper layers in the ejecta, where the density
profile is flatter, favouring extended emission above the continuum photosphere.
The velocities are also lower, so any velocity offset becomes less conspicuous.
Moreover, time-dependent ionization exacerbates this effect by making the line
optical depth more slowly varying with radius/velocity \citep{des08_2}, also favouring
extended line emission. Interestingly, it is in part the increase in the extension
of the spectrum formation region that favours the rise of polarization at the end
of the plateau phase in SNe~II-P \citep{des11_2}.\\
\indent Departures from the mean trajectory of the peak location in velocity space of emission peaks
are visible for
three models in Fig.\ 6. First, the two models with a lower/higher kinetic energies (models m15e0p6 and
m15e3p0) show smaller/larger offsets, simply reflecting the contrast in ejecta velocity 
(same envelope mass $M$, but different kinetic energy $E$).
In these, the peak blueshifts at 10\,d after explosion are  $-$2000\,\kms\ and $-$3500\,\kms\ respectively. 
Second, model m15mlt1x3, which has a higher ejecta kinetic energy to ejecta mass 
(same $E$ of 1.2\,B as the models of \citealt{des13}, but much lower H-envelope mass),
shows a markedly different trajectory for the peak-emission offset. Because the expansion rate
is much larger, the offset is large initially, but the reduced envelope mass makes the transition
to nebular phase earlier, and at such times the offset is always found to be zero in our models.
Thus, the rate of change of this offset is much larger in model  m15mlt1x3. This is an interesting 
property which is also seen in observations. Indeed, SNe with a higher $s_2$ parameter (i.e.,
$V$-band light curves with steeper decline rates) also have larger initial velocity offsets (see Fig.~\ref{figs2}).

\subsection{Correlation with other SN properties}

\begin{figure}
\includegraphics[width=8.5cm]{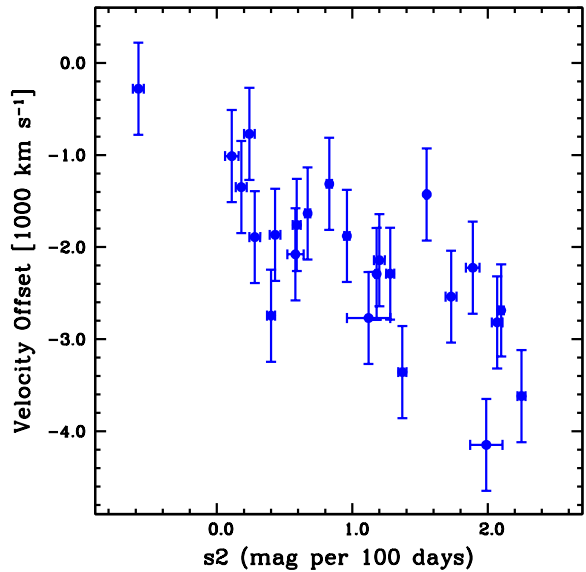}
\caption{$s_2$ (the decline rate during the `plateau') plotted against the
  velocity offset of SNe \ha\ emission peaks at 30\,d\ post explosion.\label{figs2}}
\end{figure}

\begin{figure}
\includegraphics[width=8.5cm]{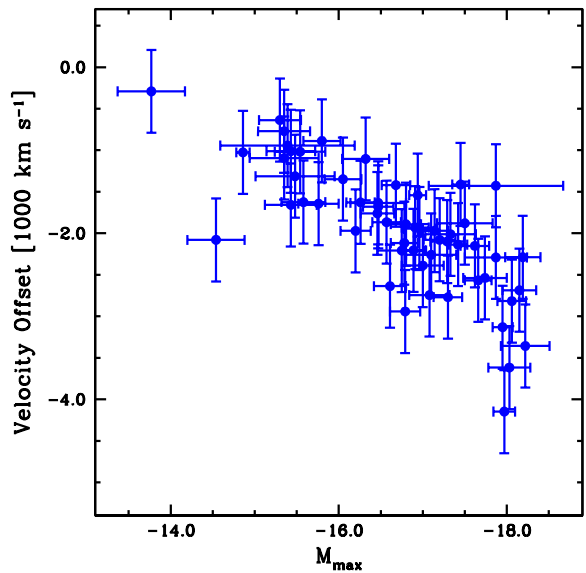}
\caption{$M_\text{max}$ (absolute magnitude at maximum light in the $V$-band) plotted against the
  velocity offset of SNe \ha\ emission peaks at 30\,d\ post explosion. \label{figmax}}
\end{figure}

While a full investigation into how observations of blue-shifted emission are related to other
SN properties is beyond the 
scope of this paper, in this Section we outline a couple of interesting correlations that 
have been found, and suggest how these may be used to further understand the nature of SNe~II.
In Fig.~\ref{figs2}  the decline rate during the `plateau', $s_2$ (which is a proxy for the 
degree to which a type II SN could be classified as `Linear': see Anderson et al. accepted), is plotted against
the SN emission peak velocity offset measured at 30\,d\ post explosion (only SNe which have both $s_1$ --the initial decline
from maximum--
and $s_2$ defined are included, see Anderson et al. accepted for justification of this). A
correlation is found in that faster declining SNe show higher velocity offsets
at the same epoch as compared to smaller $s_2$ SNe. To test the significance of this
correlation we employ a Monte Carlo application of the Pearson's test for correlations (following the 
procedure outline in Anderson et al. accepted). A Pearson's r-value of $-0.75\pm0.10$ is calculated, which, for $N= 24$ 
equates 
to a lower limit significance $P= 6\times10^{-4}$ (i.e. the probability of finding a correlation by chance).
It is interesting that in the synthetic measurements displayed in Fig.~\ref{fig_path},
the highest velocity offsets at around 30\,d\ post explosion 
are for the model (m15mlt1x3), which has a more `Linear' light-curve morphology (i.e. a high $s_2$). 
As discussed in Section~\ref{sect_res_mod}, model m15mlt1x3 is characterized by the same
ejecta kinetic energy of 1.2\,B as other models from \citet{des13}, but its ejecta mass is smaller.
This translates to a higher expansion rate and a shorter photospheric phase duration, hence
a larger \ha\ peak offset early on together with its more rapid decrease to zero.\\ 
\indent In Fig.~\ref{figmax}, SNe maximum $V$-band magnitudes are correlated against 
SN emission peak velocity offsets measured at 30\,d\ post explosion. A
strong correlation is found in
terms of brighter SNe showing higher velocity offsets (confirmed by a Monte Carlo application of 
the Pearson's test: $N=51$, $r=0.75\pm0.07$ $P<2\times10^{-8}$). The strength of this correlation is quite
remarkable given the uncertainties in measurements of  $M_\text{max}$ (problems in defining
this epoch, issues with host galaxy extinction corrections), the explosion epoch estimations
(see discussion in Anderson et al. accepted), together with those of the velocity offsets. 
Here again, our more `Linear' type II SN model m15mlt1x3 offers a promising explanation (together with the possibility of 
diversity being related to explosion energy differences).
Indeed, it exhibits a much higher luminosity than other SNe II-P models in \citet{des13},
although it has the same explosion energy. The difference here is that this energy is coupled
to a smaller progenitor envelope mass, leading to a higher energy per unit mass, a higher
luminosity early on with a larger expansion rate. Because the ejecta mass is lower, the photospheric
phase is shorter and so the luminosity decreases earlier and faster until reaching the nebular phase.
In that context, this observation would support the notion that the diversity of SNe II-P, in particular
the smooth connection to type II SNe with more `Linear' light curves 
(Anderson et al., accepted, although see \citealt{arc12} for claims of a distinct separation between
SNe~II-P and SNe~II-L), stems from a diversity in progenitor envelope mass.
This hypothesis is discussed at length in Anderson et al. (accepted) with respect to the diversity
seen in the $V$-band light-curves of 116 SNe~II. 
Of course, this does not preclude the additional diversity stemming from varying explosion energies.\\
\indent In the context of the current analysis, the diversity of the observed (and predicted) strength of blue-shifted emission 
velocities can then be understood in terms of changes in the pre-SN envelope mass and/or differences in explosion energy. As one decreases the envelope mass (or increases the explosion energy), 
ejecta velocities increase
due to the higher energy per unit mass. The strength of the blue-shifted emission velocity is then directly linked to the
ejecta velocity. This energy increase also increases the early luminosity (hence a higher $M_\text{max}$, as 
observed in Fig.~\ref{figmax}), while 
the faster expansion leads to faster declining light-curves, and hence the correlation observed
with $s_2$ in Fig.~\ref{figs2}.\\
\indent Finally, we note that the correlation presented in Fig.~\ref{figmax} implies that measurements of
the blueshift of \ha\ emission profiles could be used to predict SN absolute magnitudes. 
In this sense, such a correlation could be used as another SN~II distance indicator method 
(to complement other such methods: e.g. \citealt{ham02}). The RMS scatter of the trend found in Fig.~\ref{figmax} 
is $\sim$0.68 mag, however one may hope to further improve this with a more detailed analysis (e.g. better constraints
on host extinction, the use of colour information).
As noted in \S\ 2, the follow-up surveys which contributed to the current sample were magnitude limited
in nature. Hence, if the correlation presented in Fig.~\ref{figmax} holds, then
surveys such as the CSP will preferentially follow SNe with larger blue-shifted velocity offsets.

\section{Conclusions}
We have presented an observational analysis which shows that
blue-shifted emission peaks are seen in \textit{all} spectral sequences of
SNe~II. These blue-shifted velocity offsets are observed to be larger at early
times and evolve to being consistent with zero once SNe enter their nebular 
phase. Hence, peak-emission blueshifts are generic properties of observed SN~II spectra. 
In addition,  we demonstrate that emission-peak blueshifts are also a generic feature of 
photospheric-phase SN~II spectra computed with the non-LTE time-dependent 
radiative-transfer code \cmfgen\ and based on physical models of the progenitor
star and explosion. In fact, such blueshifts occur not just in
SNe~II, but also in SNe Ia \citep{blo06_2}, although line overlap complicates
a clear identification of such offsets. In SNe~II, the velocity offset is best seen in H$\alpha$,
because the line is strong, but also because one can track the velocity offset from being large early on
to becoming negligible when the SN turns nebular.\\
\indent In addition, we have shown that the diversity of blue-shifted velocities found
within a large sample of SNe~II, can also be linked to SN~II light-curve diversity. SNe
showing larger blueshifts are also found to be brighter objects, and have faster declining 
light-curves. We speculate that this diversity and the trends observed between different
parameters, can be most easily understood in terms of differences in proegnitor envelope masses
retained before the epoch of explosion.\\
\indent Until now, measurements, analysis and discussion of these properties for 
SNe of any kind have been scarce, despite the fact that their presence is obvious and systematic.
As our analysis and discussion show, these features and their evolution  systematically change between SNe, 
and offer additional leverage to identify the origin of the diversity in SN~II observations, progenitors,
and explosion.

\section*{Acknowledgments}
We thank the annonymous referee for their useful suggestions.
This paper is based on observations obtained at the Gemini Observatory, which is operated by the 
Association of Universities for Research in Astronomy, Inc., under a cooperative agreement 
with the NSF on behalf of the Gemini partnership: the National Science Foundation 
(United States), the National Research Council (Canada), CONICYT (Chile), the Australian 
Research Council (Australia), Minist\'{e}rio da Ci\^{e}ncia, Tecnologia e Inova\c{c}\~{a}o 
(Brazil) and Ministerio de Ciencia, Tecnolog\'{i}a e Innovaci\'{o}n Productiva (Argentina)
(Gemini Program GS-2008B−Q−56).
This paper includes data gathered with the 6.5 meter Magellan Telescopes located at Las Campanas Observatory, Chile. 
Based on observations made with ESO telescopes at the La Silla Paranal Observatory under programme
under programmes: 076.A-0156,
078.D-0048, 080.A-0516, and 082.A-0526.
J. P. Anderson acknowledges support by CONICYT through
FONDECYT grant 3110142, and by the Millennium Center for
Supernova Science (P10-064-F), with input from `Fondo de
Innovación para la Competitividad, del Ministerio de
Economía, Fomento y Turismo de Chile'. 
L. Dessart acknowledges financial support from the European Community through an
International Re-integration Grant, under grant number PIRG04-GA-2008-239184,
and from ``Agence Nationale de la Recherche" grant ANR-2011-Blanc-SIMI-5-6-007-01.
The work of the CSP has been supported by the 
National Science Foundation 
under grants AST0306969, AST0607438, and AST1008343.
M.~H. and C.~G. acknowledge support by projects IC120009 ``Millennium Institute
of Astrophysics (MAS)" and P10-064-F "Millennium Center for Supernova
Science" of the Iniciativa Científica Milenio del Ministerio Economía,
Fomento y Turismo de Chile.
M.~D.~S. gratefully acknowledges generous support 
provided by the Danish Agency for Science and Technology and Innovation  
realized through a Sapere Aude Level 2 grant.
This research has made use of the NASA/IPAC Extragalactic Database (NED) 
which is operated by the Jet 
Propulsion Laboratory, California
Institute of Technology, under contract with the National Aeronautics.

\bibliographystyle{mn2e}
\bibliography{Reference}

\begin{thebibliography}{}

\bibitem[\protect\citeauthoryear{{Arcavi} et~al.,}{{Arcavi}
  et~al.}{2012}]{arc12}
{Arcavi} I.,  et~al., 2012, \apjl, 756, L30

\bibitem[\protect\citeauthoryear{{Blondin} et~al.,}{{Blondin}
  et~al.}{2006}]{blo06_2}
{Blondin} S.,  et~al., 2006, \aj, 131, 1648

\bibitem[\protect\citeauthoryear{{Bose} et~al.,}{{Bose}  et~al.}{2013}]{bos13}
{Bose} S.,  et~al., 2013, \mnras, 433, 1871

\bibitem[\protect\citeauthoryear{{Cappellaro}, {Danziger}, {della Valle},
  {Gouiffes} \& {Turatto}}{{Cappellaro} et~al.}{1995}]{cap95}
{Cappellaro} E.,  {Danziger} I.~J.,  {della Valle} M.,  {Gouiffes} C.,
  {Turatto} M.,  1995, \aap, 293, 723

\bibitem[\protect\citeauthoryear{{Castor}}{{Castor}}{1970}]{cas70}
{Castor} J.~I.,  1970, \mnras, 149, 111

\bibitem[\protect\citeauthoryear{{Chevalier}}{{Chevalier}}{1976}]{che76}
{Chevalier} R.~A.,  1976, \apj, 207, 872

\bibitem[\protect\citeauthoryear{{Chugai}}{{Chugai}}{1988}]{chu88}
{Chugai} N.~N.,  1988, Soviet Astronomy Letters, 14, 334

\bibitem[\protect\citeauthoryear{{Dessart} et~al.,}{{Dessart}
  et~al.}{2008}]{des08}
{Dessart} L.,  et~al., 2008, \apj, 675, 644

\bibitem[\protect\citeauthoryear{{Dessart} \& {Hillier}}{{Dessart} \&
  {Hillier}}{2005a}]{des05}
{Dessart} L.,  {Hillier} D.~J.,  2005a, \aap, 439, 671

\bibitem[\protect\citeauthoryear{{Dessart} \& {Hillier}}{{Dessart} \&
  {Hillier}}{2005b}]{des05_2}
{Dessart} L.,  {Hillier} D.~J.,  2005b, \aap, 437, 667

\bibitem[\protect\citeauthoryear{{Dessart} \& {Hillier}}{{Dessart} \&
  {Hillier}}{2008}]{des08_2}
{Dessart} L.,  {Hillier} D.~J.,  2008, \mnras, 383, 57

\bibitem[\protect\citeauthoryear{{Dessart} \& {Hillier}}{{Dessart} \&
  {Hillier}}{2010}]{des10_2}
{Dessart} L.,  {Hillier} D.~J.,  2010, \mnras, 405, 2141

\bibitem[\protect\citeauthoryear{{Dessart} \& {Hillier}}{{Dessart} \&
  {Hillier}}{2011}]{des11_2}
{Dessart} L.,  {Hillier} D.~J.,  2011, \mnras, 415, 3497

\bibitem[\protect\citeauthoryear{{Dessart}, {Hillier}, {Waldman} \&
  {Livne}}{{Dessart} et~al.}{2013}]{des13}
{Dessart} L.,  {Hillier} D.~J.,  {Waldman} R.,    {Livne} E.,  2013, \mnras,
  433, 1745

\bibitem[\protect\citeauthoryear{{Dessart}, {Livne} \& {Waldman}}{{Dessart}
  et~al.}{2010}]{des10}
{Dessart} L.,  {Livne} E.,    {Waldman} R.,  2010, \mnras, 408, 827

\bibitem[\protect\citeauthoryear{{Elmhamdi}, {Danziger}, {Chugai},
  {Pastorello}, {Turatto}, {Cappellaro}, {Altavilla}, {Benetti}, {Patat} \&
  {Salvo}}{{Elmhamdi} et~al.}{2003}]{elm03_2}
{Elmhamdi} A.,  {Danziger} I.~J.,  {Chugai} N.,  {Pastorello} A.,  {Turatto}
  M.,  {Cappellaro} E.,  {Altavilla} G.,  {Benetti} S.,  {Patat} F.,    {Salvo}
  M.,  2003, \mnras, 338, 939

\bibitem[\protect\citeauthoryear{{Filippenko}}{{Filippenko}}{1997}]{fil97}
{Filippenko} A.~V.,  1997, \araa, 35, 309

\bibitem[\protect\citeauthoryear{{Folatelli} et~al.,}{{Folatelli}
  et~al.}{2013}]{fol13}
{Folatelli} G.,  et~al., 2013, \apj, 773, 53

\bibitem[\protect\citeauthoryear{{Fraser} et~al.,}{{Fraser}
  et~al.}{2011}]{fra09}
{Fraser} M.,  et~al., 2011, \mnras, 417, 1417

\bibitem[\protect\citeauthoryear{{Hamuy} et~al.,}{{Hamuy}
  et~al.}{2001}]{ham01}
{Hamuy} M.,  et~al., 2001, \apj, 558, 615

\bibitem[\protect\citeauthoryear{{Hamuy} et~al.,}{{Hamuy}
  et~al.}{2006}]{ham06}
{Hamuy} M.,  et~al., 2006, \pasp, 118, 2

\bibitem[\protect\citeauthoryear{{Hamuy} \& {Pinto}}{{Hamuy} \&
  {Pinto}}{2002}]{ham02}
{Hamuy} M.,  {Pinto} P.~A.,  2002, \apjl, 566, L63

\bibitem[\protect\citeauthoryear{{Hillier} \& {Dessart}}{{Hillier} \&
  {Dessart}}{2012}]{hil12}
{Hillier} D.~J.,  {Dessart} L.,  2012, \mnras, 424, 252

\bibitem[\protect\citeauthoryear{{Leonard} et~al.,}{{Leonard}
  et~al.}{2002a}]{leo02_2}
{Leonard} D.~C.,  et~al., 2002a, \aj, 124, 2490

\bibitem[\protect\citeauthoryear{{Leonard} et~al.,}{{Leonard}
  et~al.}{2002b}]{leo02}
{Leonard} D.~C.,  et~al., 2002b, \pasp, 114, 35

\bibitem[\protect\citeauthoryear{{Menzies} et~al.,}{{Menzies}
  et~al.}{1987}]{men87}
{Menzies} J.~W.,  et~al., 1987, \mnras, 227, 39P

\bibitem[\protect\citeauthoryear{{Minkowski}}{{Minkowski}}{1941}]{min41}
{Minkowski} R.,  1941, \pasp, 53, 224

\bibitem[\protect\citeauthoryear{{Quimby}, {Wheeler}, {H{\"o}flich}, {Akerlof},
  {Brown} \& {Rykoff}}{{Quimby} et~al.}{2007}]{qui07}
{Quimby} R.~M.,  {Wheeler} J.~C.,  {H{\"o}flich} P.,  {Akerlof} C.~W.,  {Brown}
  P.~J.,    {Rykoff} E.~S.,  2007, \apj, 666, 1093

\bibitem[\protect\citeauthoryear{{Sobolev}}{{Sobolev}}{1960}]{sob60}
{Sobolev} V.~V.,  1960, {Moving envelopes of stars}

\bibitem[\protect\citeauthoryear{{Turatto}, {Cappellaro}, {Benetti} \&
  {Danziger}}{{Turatto} et~al.}{1993}]{tur93}
{Turatto} M.,  {Cappellaro} E.,  {Benetti} S.,    {Danziger} I.~J.,  1993,
  \mnras, 265, 471

\end{thebibliography}

\label{lastpage}

\end{document}